\documentstyle[epsfig,my,cite,amssymb]{aipproc}

\def\etjet{E_T^{jet}}
\def\etajet{\eta^{jet}}

\def\qg2{$\q2>125$~\g2}

\def\q2{Q^2}
\def\pb1{pb$^{-1}$}
\def\gp{\gamma p}
\def\eg{e\gamma}
\def\ga{\gamma\gamma}
\def\g2{GeV$^2$}

\def\F2g{F_2^{\gamma}}
\def\f2gv{F_2^{\gamma^*}}

\def\rr1{R=1.0}
\def\r7{R=0.7}
\def\R71{R=0.7\ {\rm and}\ 1.0}

\def\m3j{M^{3J}}

\def\kt{k_T}

\def\lq2{\log_{10}(\q2)}

\def\xo{x_{\gamma}^{OBS}}

\def\Journal#1#2#3#4{{#1} {#2} (#3) #4}

\def\NPB{{\em Nucl. Phys.} {\bf B}}
\def\PLB{{\em Phys. Lett.}  {\bf B}}

\def\PRD{{\em Phys. Rev.} {\bf D}}

\def\ZPC{{\em Z. Phys.} {\bf C}}

\def\EPC{{\em Eur. Phys. Jour.} {\bf C}}
\def\EPD{{\em Eur. Phys. Jour. direct} {\bf C}}
\def\CPC{{\em Comp. Phys. Comm.}}

\def\colab#1{{#1 Collaboration}}

\def\conf#1#2#3#4{{paper #1}, {submitted to the #2}, {#3} {(#4)}}

\def\talk#1#2#3#4#5{{Talk given at the {\it #1}}, {#2}, {#3}, {#4}, {#5}.}

\def\presub#1{{submitted to #1}}

\def\etal{{\it et al}}

\def\sxo{d\sigma/d\xo}

\def\z0{Z^{\circ}}

\def\oalphas2{O(\alpha\alpha_S^2)}

\def\p2{P^2}

\def\figdir{figures/}

\begin{document}

\title{The Structure of the Virtual Photon\footnotemark[1]}

\author{Claudia Glasman$^*$\\
representing the ZEUS Collaboration}

\address{$^*$Department of Physics and Astronomy, Kelvin Building,\\
University of Glasgow, Glasgow, G12 8QQ, UK}

\maketitle

\begin{abstract}
Measurements of dijet cross sections for virtual photons are presented as a
function of $\xo$, the fraction of the virtual photon energy invested in the
production of the dijet system, using the ZEUS detector. Comparisons to QCD
predictions show that a resolved photon component is needed to describe the
data up to values of the photon virtuality comparable to the scale of the
interaction.
\end{abstract}

\fntext{}{\talk{International Conference on the Structure and Interactions of
the Photon {\rm (PHOTON 2000)}}{Ambleside}{UK}{August $26^{th}-31^{st}$}{2000}}

\section*{INTRODUCTION}

It has been long established that real photons have a partonic structure from
measurements of the photon structure function $\F2g$ in $\eg$ interactions
\cite{eg} and observation of resolved photon processes in $\gp$ interactions
\cite{gp} and of single- and double-resolved processes in $\ga$ interactions
\cite{gg}. Thus, it is natural to expect that virtual photons also
have a partonic structure \cite{SaS,DG,theory,predictions}. QCD predicts that
the parton densities of virtual photons become logarithmically suppressed as
the virtuality of the probed photon $\p2$ increases for fixed $\mu^2$, the
scale of the interaction; $\mu^2$ is usually taken as $\q2$ ($\q2$ is the
virtuality of the probing photon) in deep inelastic scattering (DIS) $\eg$ and
the jet transverse energy ($\etjet$) in jet production.

\begin{figure*}
\begin{center}
\setlength{\unitlength}{1.0cm}
\begin{picture} (5.0,4.0)
\put (-3.0,0.2){\epsfig{figure=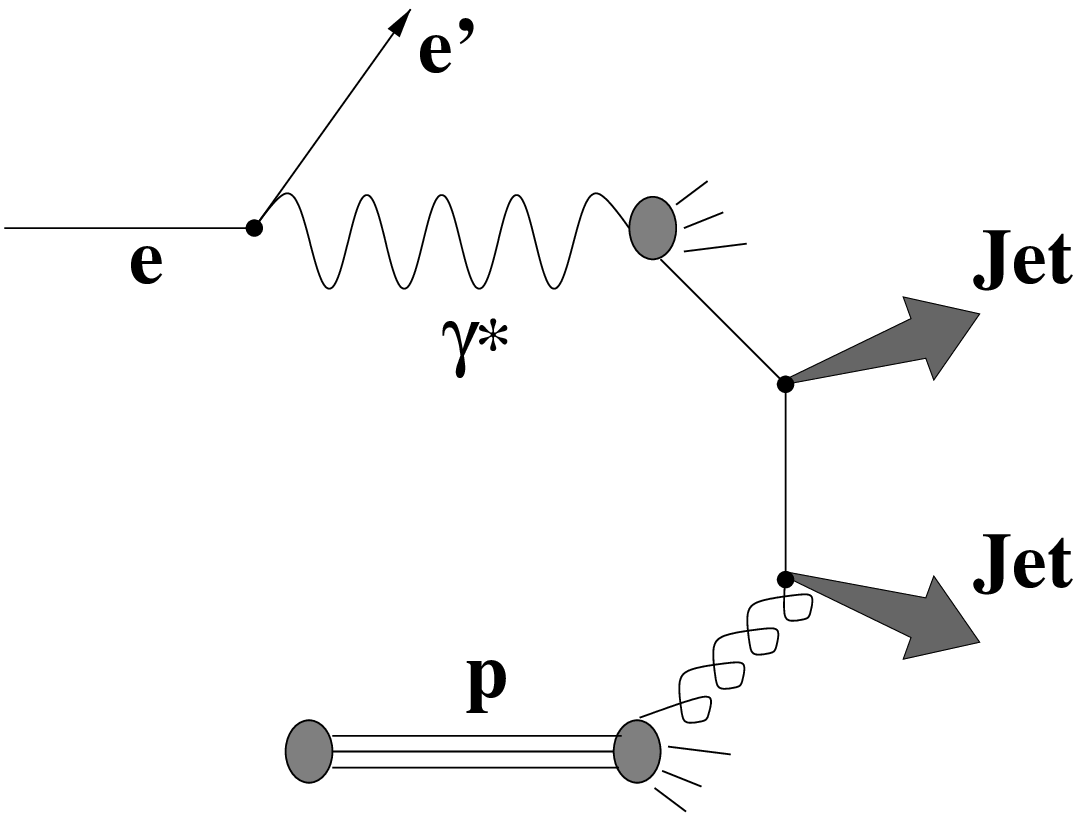,width=5cm}}
\put (4.0,0.2){\epsfig{figure=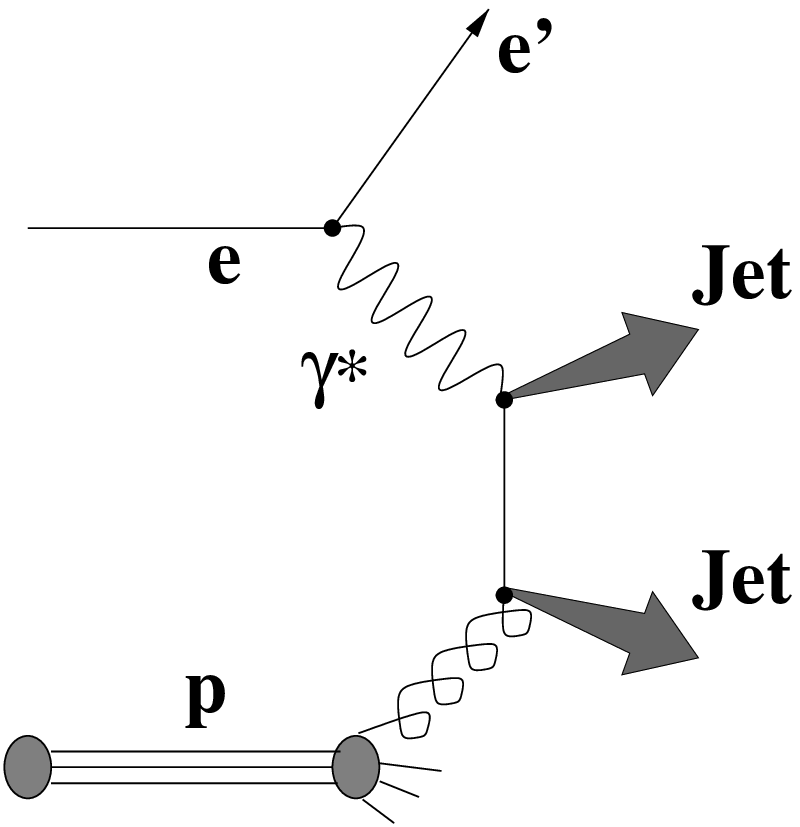,width=4cm}}
\put (-0.2,-0.3){\small (a)}
\put (5.5,-0.3){\small (b)}
\end{picture}
\end{center}
\caption{\label{fig1} (a) Resolved and (b) direct virtual-photon processes.}
\end{figure*}

The virtual photon structure function $\f2gv$ at leading order (LO) in QCD is
given by

$$\f2gv(x_{\gamma^*},\p2,\mu^2_{F_{\gamma^*}})=\displaystyle\sum_qx_{\gamma^*}\ e_q^2\ [f_{q/\gamma^*}(x_{\gamma^*},\p2,\mu^2_{F_{\gamma^*}})+f_{\bar q/\gamma^*}(x_{\gamma^*},\p2,\mu^2_{F_{\gamma^*}})],$$
where $x_{\gamma^*}$ is the fraction of the virtual photon momentum taken by
the interacting parton, $\mu^2_{F_{\gamma^*}}$ is the virtual photon
fragmentation scale and $e_q$ is the quark charge. The virtual photon
quark densities $f_{q/\gamma^*}$ contain two terms, as in the case of
the real photon,

$$f_{i/\gamma^*}(x_{\gamma^*},\p2,\mu^2_{F_{\gamma^*}})=f^{\rm had}_{i/\gamma^*}(x_{\gamma^*},\p2,\mu^2_{F_{\gamma^*}})+
f^{\rm anom}_{i/\gamma^*}(x_{\gamma^*},\p2,\mu^2_{F_{\gamma^*}}),$$
one term associated to the non-perturbative hadronic component
($f^{\rm had}$) and a term $f^{\rm anom}$, unique to the photon, which
expresses the direct coupling of the photon to a $q\bar q$ pair, calculable in
perturbative QCD (pQCD). The hadronic component is the one expected to
decrease as $\p2$ increases. The first measurements of the structure function
of the virtual photon were done by PLUTO \cite{pluto}. Only recently new
measurements from LEP have become available \cite{l3}.

At HERA, the virtual photon structure is studied in jet production mediated
by virtual photons. In LO QCD, two processes are expected to contribute to the
jet cross section: resolved processes (figure \ref{fig1}a) in which the
virtual photon interacts with the proton via its hadronic component giving two
jets in the final state, and the direct processes (figure \ref{fig1}b) in
which the virtual photon interacts as a point-like particle with a parton from
the proton.

The dijet production cross sections in LO QCD for direct and resolved
processes are given by

$$\sigma^{LO,ep}_{D} = \int d\Omega\ f_{\gamma^* /e}(y,\p2)\ 
f_{j/p}(x_p,\mu^2_{F_p})\ d\sigma(\gamma^* j\rightarrow {\rm jet}\ {\rm jet})$$
and
$$\sigma^{LO,ep}_{R} = \int d\Omega\ f_{\gamma^* /e}(y,\p2)\ 
f_{i/\gamma^*}(x_{\gamma^*},\p2,\mu^2_{F_{\gamma^*}})\ f_{j/p}(x_p,\mu^2_{F_p})\
d\sigma(ij\rightarrow {\rm jet}\ {\rm jet}),$$
where the integrals are performed over the phase space represented by
``$d\Omega$''; $f_{\gamma^*/e}(y,\p2)$ is the flux of virtual photons in the
positron and $y$ is the fraction of the positron energy taken by the virtual
photon; $f_{j/p}(x_p,\mu^2_{F_p})$ are the parton densities in the proton and
$x_p$ is the fraction of the proton momentum taken by parton $j$;
$\mu^2_{F_p}$ is the proton fragmentation scale; and
$d\sigma(\gamma^*(i)j\rightarrow {\rm jet}\ {\rm jet})$ is the subprocess
cross section, calculable in pQCD. In the case of resolved processes, there is
an additional ingredient: the parton densities in the virtual photon
$f_{i/\gamma^*}(x_{\gamma^*},\p2,\mu^2_{F_{\gamma^*}})$, for which up to the
present there is only very little experimental information. This allows the
study of the virtual photon structure by measuring jet cross sections. 

The contribution to the dijet cross section from resolved processes is expected
to decrease relative to the contribution from the direct processes as
$\p2\rightarrow\mu^2$; this means that the partonic content of the virtual
photon becomes suppressed as $\p2$ increases. Experimentally, resolved and
direct processes are separated by using the variable $\xo$,

\begin{equation}
\xo={1\over 2yE_e}({E_T^{jet1}e^{-\eta^{jet1}}+E_T^{jet2}e^{-\eta^{jet2}}}),
\label{eqone}
\end{equation}
where $\eta^{jet1,2}$ is the jet pseudorapidity and $E_e$ is the positron
beam energy. $\xo$ measures the fraction of the virtual photon energy invested
in the production of the dijet system and it is well defined at all orders in
pQCD. For resolved (direct) processes, $\xo$ takes low (high) values.

The measurements presented here were performed with the ZEUS detector at HERA.
During 1995 to 1997 HERA operated with positrons of energy $E_e=27.5$~GeV
and protons of energy $E_p=820$~GeV.

\section*{DIJET CROSS SECTIONS IN THE LABORATORY FRAME}

\begin{figure*}
\begin{center}
\setlength{\unitlength}{1.0cm}
\begin{picture} (10.0,10.0)
\put (0.0,1.2){\epsfig{figure=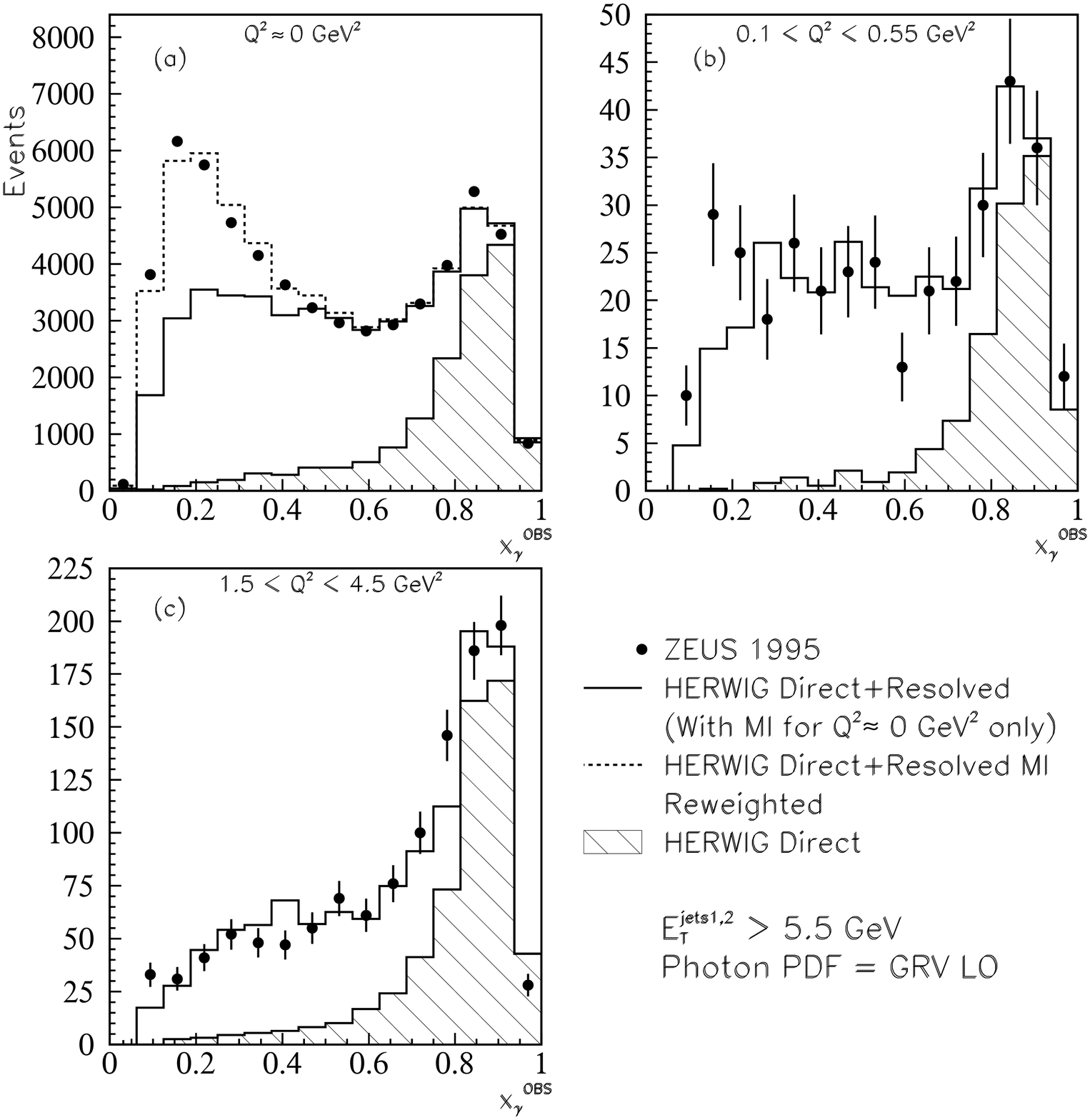,width=10cm}}
\put (7.15,4.5){\epsfig{figure=\figdir white.eps,width=0.26cm,height=0.31cm}}
\put (2.0,10.4){\epsfig{figure=\figdir white.eps,width=0.29cm,height=0.35cm}}
\put (6.9,10.4){\epsfig{figure=\figdir white.eps,width=0.29cm,height=0.35cm}}
\put (2.45,5.66){\epsfig{figure=\figdir white.eps,width=0.29cm,height=0.35cm}}
\put (-0.7,-2.15){\epsfig{figure=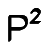,width=16cm}}
\put (-4.8,4.66){\epsfig{figure=\figdir letters1.eps,width=14cm}}
\put (0.1,4.66){\epsfig{figure=\figdir letters1.eps,width=14cm}}
\put (-4.35,-0.08){\epsfig{figure=\figdir letters1.eps,width=14cm}}
\end{picture}
\end{center}
\vspace{-1.5cm}
\caption{\label{fig2} 
The uncorrected $\xo$ distribution (black dots). Monte Carlo simulations using
HERWIG are shown for comparison.}
\end{figure*}

The $\xo$ distribution displays a high sensitivity to the virtual photon
structure. Figure \ref{fig2} shows the $\xo$ distribution for a sample of
dijet events in three regions of $\p2$. In all the $\p2$ regions studied, a
significant resolved contribution is needed in the Monte Carlo to
describe the data for $\xo<0.75$. Direct processes alone (the shaded area in
figure \ref{fig2}) cannot describe the data \cite{lab}.

\begin{figure*}
\begin{center}
\setlength{\unitlength}{1.0cm}
\begin{picture} (10.0,10.0)
\put (0.0,1.2){\epsfig{figure=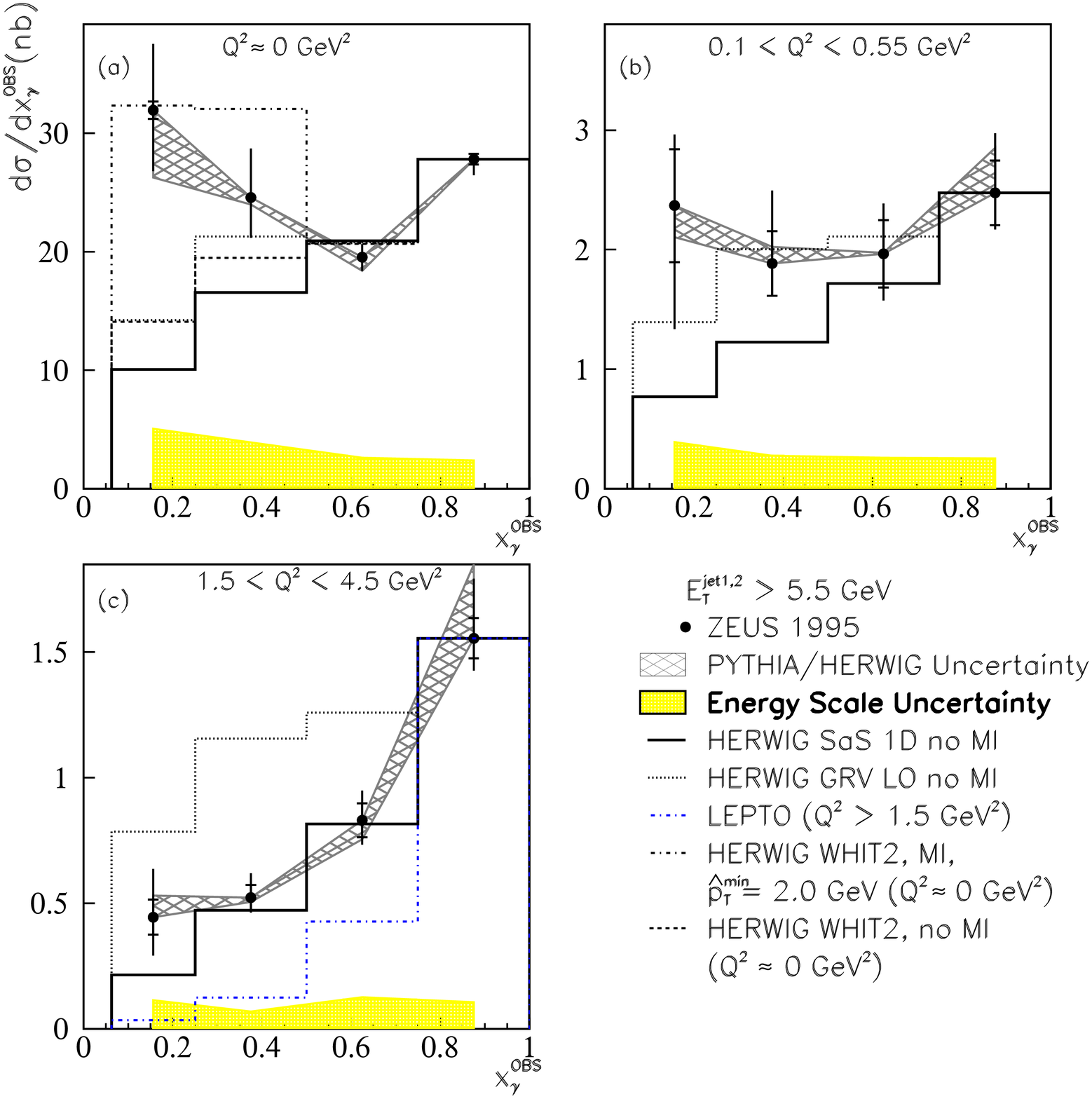,width=10cm}}
\put (7.17,3.7){\epsfig{figure=\figdir white.eps,width=0.26cm,height=0.31cm}}
\put (7.94,3.0){\epsfig{figure=\figdir white.eps,width=0.26cm,height=0.31cm}}
\put (6.35,2.35){\epsfig{figure=\figdir white.eps,width=0.26cm,height=0.31cm}}
\put (1.9,10.4){\epsfig{figure=\figdir white.eps,width=0.29cm,height=0.35cm}}
\put (6.9,10.4){\epsfig{figure=\figdir white.eps,width=0.29cm,height=0.35cm}}
\put (2.35,5.66){\epsfig{figure=\figdir white.eps,width=0.29cm,height=0.35cm}}
\put (-0.6,-3.0){\epsfig{figure=\figdir letters1.eps,width=16cm}}
\put (0.1,-3.65){\epsfig{figure=\figdir letters1.eps,width=16cm}}
\put (-1.45,-4.35){\epsfig{figure=\figdir letters1.eps,width=16cm}}
\put (-4.9,4.66){\epsfig{figure=\figdir letters1.eps,width=14cm}}
\put (0.1,4.66){\epsfig{figure=\figdir letters1.eps,width=14cm}}
\put (-4.45,-0.08){\epsfig{figure=\figdir letters1.eps,width=14cm}}
\end{picture}
\end{center}
\vspace{-1.5cm}
\caption{\label{fig3} 
The measured dijet cross section $\sxo$ (black dots) in the laboratory frame.
The inner error bars represent the statistical errors of the data, and the
outer error bars show the statistical errors and uncorrelated systematic
uncertainties added in quadrature. The shaded band displays the uncertainty
due to the absolute energy scale of the jets and the hatched band displays the
uncertainty due to the modelling of the jet fragmentation. Monte Carlo
calculations using HERWIG and LEPTO are shown for comparison.}
\end{figure*}

Dijet cross sections as a function of $\xo$ have been measured for jets
found using the longitudinally invariant $\kt$ cluster algorithm in the
inclusive mode \cite{kt}. The measurements have been performed in the
kinematic region given by $0.2<y<0.55$ and in three regions of $\p2$:
$\p2\approx 0$ (the photoproduction regime in which the scattered positron is
lost in the beam pipe, figure \ref{fig3}a), $0.1<\p2<0.55$ \g2\ (the
intermediate $\p2$ region, in which the scattered positron is detected in the
beam pipe calorimeter, figure \ref{fig3}b) and $1.5<\p2<4.5$ \g2\ (the low
$\p2$ DIS regime, in which the scattered positron is detected in the
uranium-scintillator calorimeter, figure \ref{fig3}c). The events are
required to have at least two jets with $\etjet>5.5$ GeV and
$-1.125<\etajet<2.2$. The data show that the shape of the measured cross
section changes markedly with increasing $\p2$: the cross section for low
$\xo$ values decreases faster than at high $\xo$ values.

The measurements have been compared to several Monte Carlo models using
various parametrisations of the photon parton densities (PDFs). The GRV
\cite{grv} and WHIT2 \cite{whit2} photon PDFs have been extracted for real
photons and have no $\p2$ dependence. On the other hand, the SaS 1D
\cite{SaS} photon PDFs consist of two components with different $\p2$
dependence: the hadronic part decreases as
$\approx (m_{\rho}^2/(m_{\rho}^2+\p2))^2$ with increasing $\p2$, while the
$\p2$ dependence of the anomalous component goes like $\sim\log(\mu^2/\p2)$.
The predictions of HERWIG \cite{herwig} using the GRV PDFs do not describe
the data, as expected since they have no $\p2$ dependence. The
prediction based on WHIT2 provides a reasonable description of the data in the
photoproduction and in the intermediate $\p2$ regimes. The predictions using
SaS 1D agree with the data at high $\xo$ in all regimes. The DIS Monte Carlo
LEPTO \cite{lepto} is compared to the data in the DIS regime. It
underestimates the data at low $\xo$, i.e. parton shower effects are not the
sole contribution to this region of $\xo$ and $\p2$.

\begin{figure*}
\begin{center}
\setlength{\unitlength}{1.0cm}
\begin{picture} (10.0,10.0)
\put (0.0,1.2){\epsfig{figure=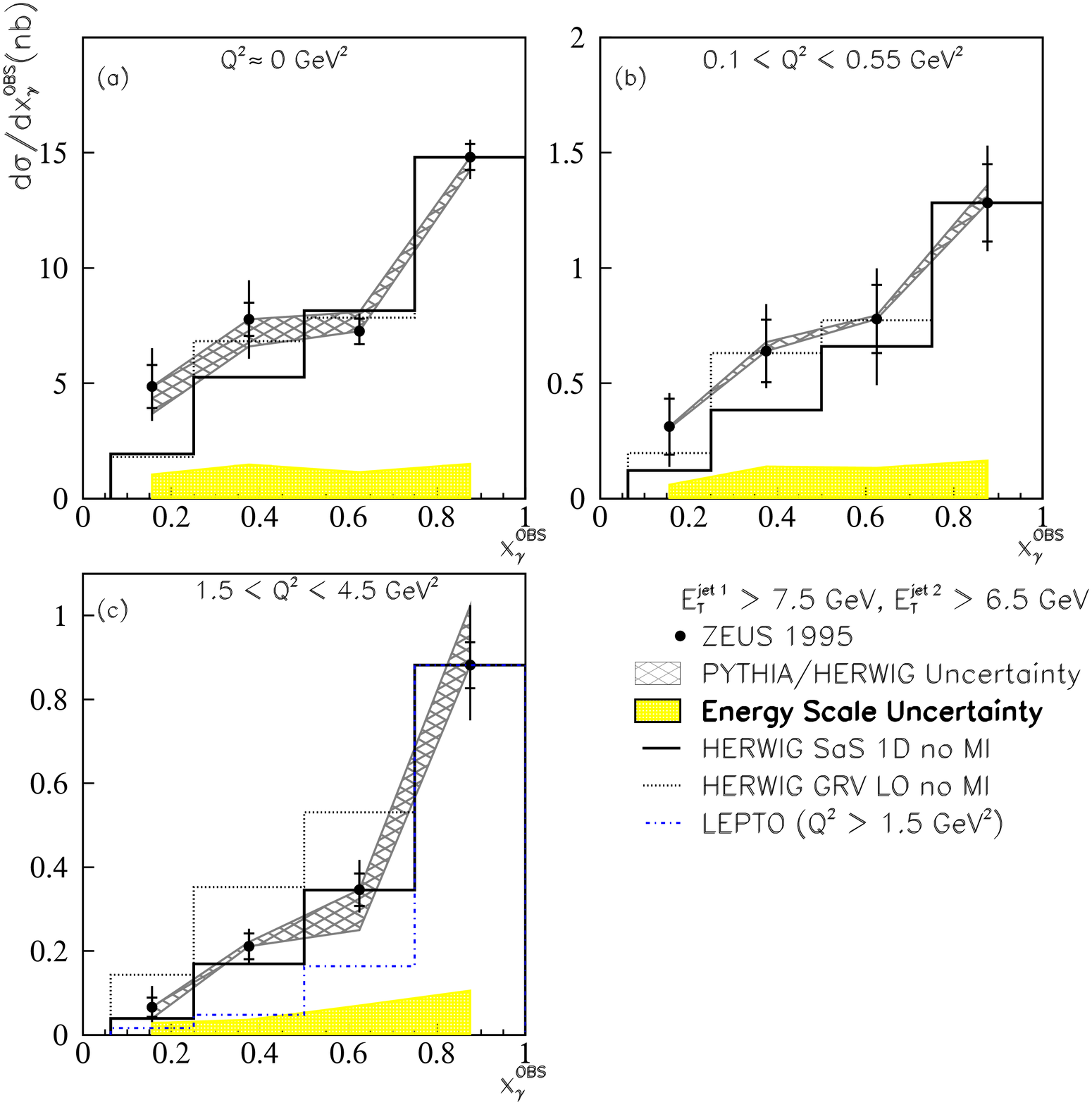,width=10cm}}
\put (7.17,3.7){\epsfig{figure=\figdir white.eps,width=0.26cm,height=0.31cm}}
\put (1.9,10.4){\epsfig{figure=\figdir white.eps,width=0.29cm,height=0.35cm}}
\put (6.9,10.4){\epsfig{figure=\figdir white.eps,width=0.29cm,height=0.35cm}}
\put (2.35,5.66){\epsfig{figure=\figdir white.eps,width=0.29cm,height=0.35cm}}
\put (-0.65,-3.0){\epsfig{figure=\figdir letters1.eps,width=16cm}}
\put (-4.9,4.66){\epsfig{figure=\figdir letters1.eps,width=14cm}}
\put (0.1,4.66){\epsfig{figure=\figdir letters1.eps,width=14cm}}
\put (-4.45,-0.08){\epsfig{figure=\figdir letters1.eps,width=14cm}}
\end{picture}
\end{center}
\vspace{-1.5cm}
\caption{\label{fig4} 
The measured dijet cross section $\sxo$ (black dots) in the laboratory frame.
Monte Carlo calculations using HERWIG and LEPTO are shown for comparison.}
\end{figure*}

The low $\etjet$ region is affected by contributions from the presence of a
possible underlying event, which may mask the $\p2$ evolution of the virtual
photon. Therefore, measurements of the dijet cross sections have been
performed at higher transverse energies. For these measurements, two jets were
required with $E_T^{jet,1(2)}>7.5(6.5)$ GeV and $-1.125<\eta^{jet1,2}<1.875$.
The cross sections are shown in figure \ref{fig4}. At low $\xo$ the cross
sections are also observed to decrease faster with $\p2$ than at high $\xo$.
The predictions from HERWIG based on the GRV PDFs are in good agreement with
the data in the two lowest $\p2$ ranges, but fail at higher $\p2$. On the
other hand, the predictions based on SaS 1D agree with the data at high $\p2$
but underestimate the data at low and intermediate $\p2$ at low $\xo$.
The predictions from LEPTO at high $\p2$ underestimate the data.

\begin{figure*}
\begin{center}
\setlength{\unitlength}{1.0cm}
\begin{picture} (10.0,10.0)
\put (0.0,1.2){\epsfig{figure=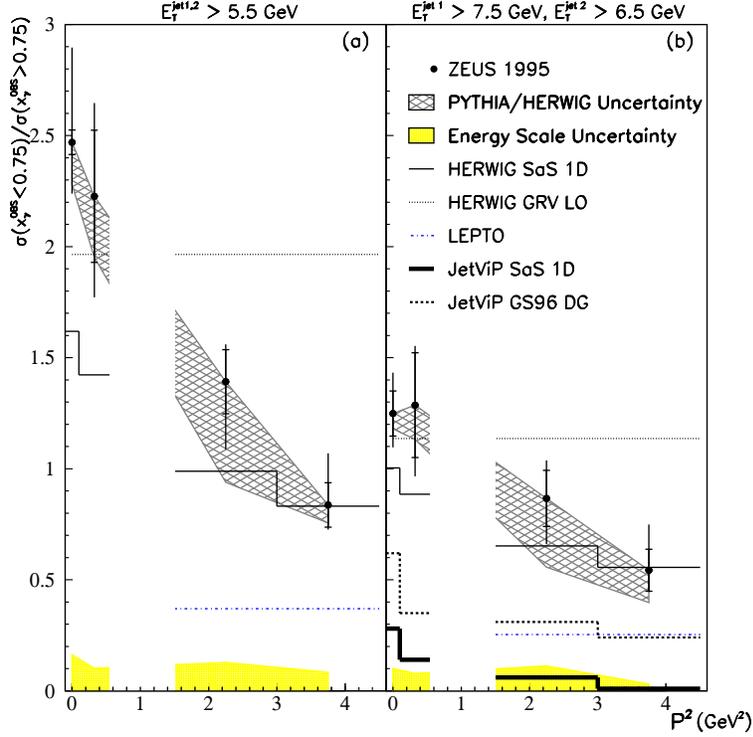,width=10cm}}
\put (8.8,1.36){\epsfig{figure=\figdir white.eps,width=0.26cm,height=0.31cm}}
\put (0.95,-5.29){\epsfig{figure=\figdir letters1.eps,width=16cm}}
\end{picture}
\end{center}
\vspace{-1.5cm}
\caption{\label{fig5}
The ratio of dijet cross sections, $\sigma(\xo<0.75)/\sigma(\xo>0.75)$, 
as a function of $\p2$ (black dots). Monte Carlo (HERWIG and LEPTO) and QCD
(JetVip) calculations are shown for comparison.}
\end{figure*}

The $\p2$ evolution of the virtual photon structure was studied further by
measuring the ratio of the dijet cross section for $\xo<0.75$ (enriched in
resolved processes) to the dijet cross section for $\xo>0.75$ (enriched in
direct processes). The ratio (see figure \ref{fig5}) falls steeply with $\p2$
which may be interpreted as the suppression of the resolved photon component
as the virtuality of the photon increases. The prediction of HERWIG based on
the GRV PDFs is constant, as expected from a photon structure without $\p2$
dependence. The prediction of SaS 1D decreases with $\p2$ but lies below the
data at low $\p2$ and the prediction from LEPTO shows that the contribution to
the ratio from parton shower effects alone is not enough to explain the $\p2$
dependence of the data. QCD predictions have been calculated using the program
JetViP \cite{jetvip} and different parametrisations of the photon PDFs. The
predictions show sensitivity to the choice of PDF but lie well below the data.
Hadronisation corrections ($\sim 20-30\%$) are insufficient to explain the
discrepancy.

\section*{DIJET CROSS SECTIONS IN THE $\gamma^*p$ FRAME}

For $\p2\gg 0$, the definition of $\xo$ (equation \ref{eqone}) is still
valid. However, the photon remnant may have high transverse energy in the
laboratory frame and be considered as a jet. The solution to avoid this
problem is to transform to the $\gamma^*p$ frame. In such a frame the photon
remnant has very low transverse energy and will not be mistaken for a jet
emanating from the hard interaction.

\begin{figure*}
\begin{center}
\setlength{\unitlength}{1.0cm}
\begin{picture} (10.0,10.0)
\put (-3.0,2.0){\epsfig{figure=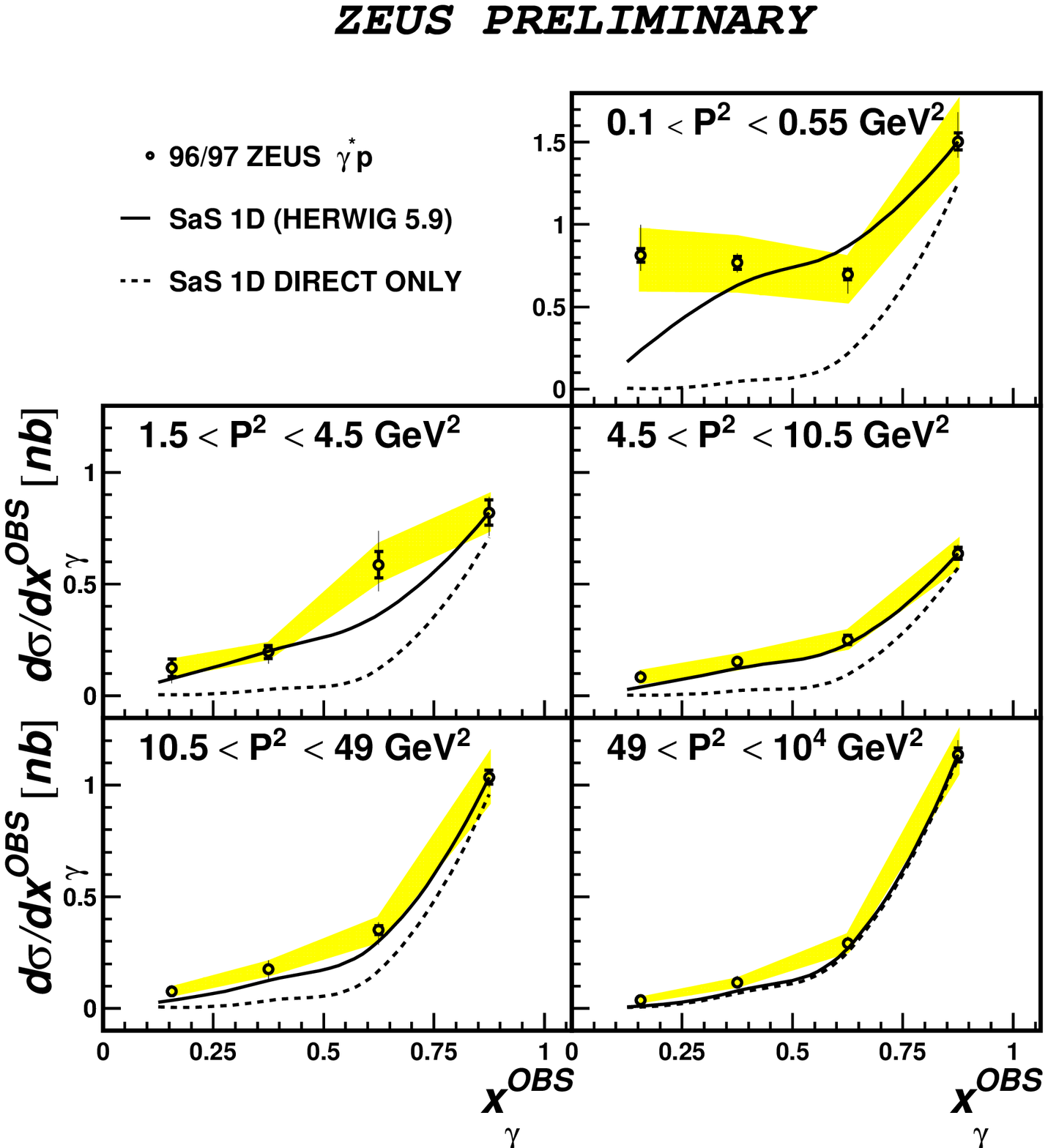,width=8cm}}
\put (5.0,2.0){\epsfig{figure=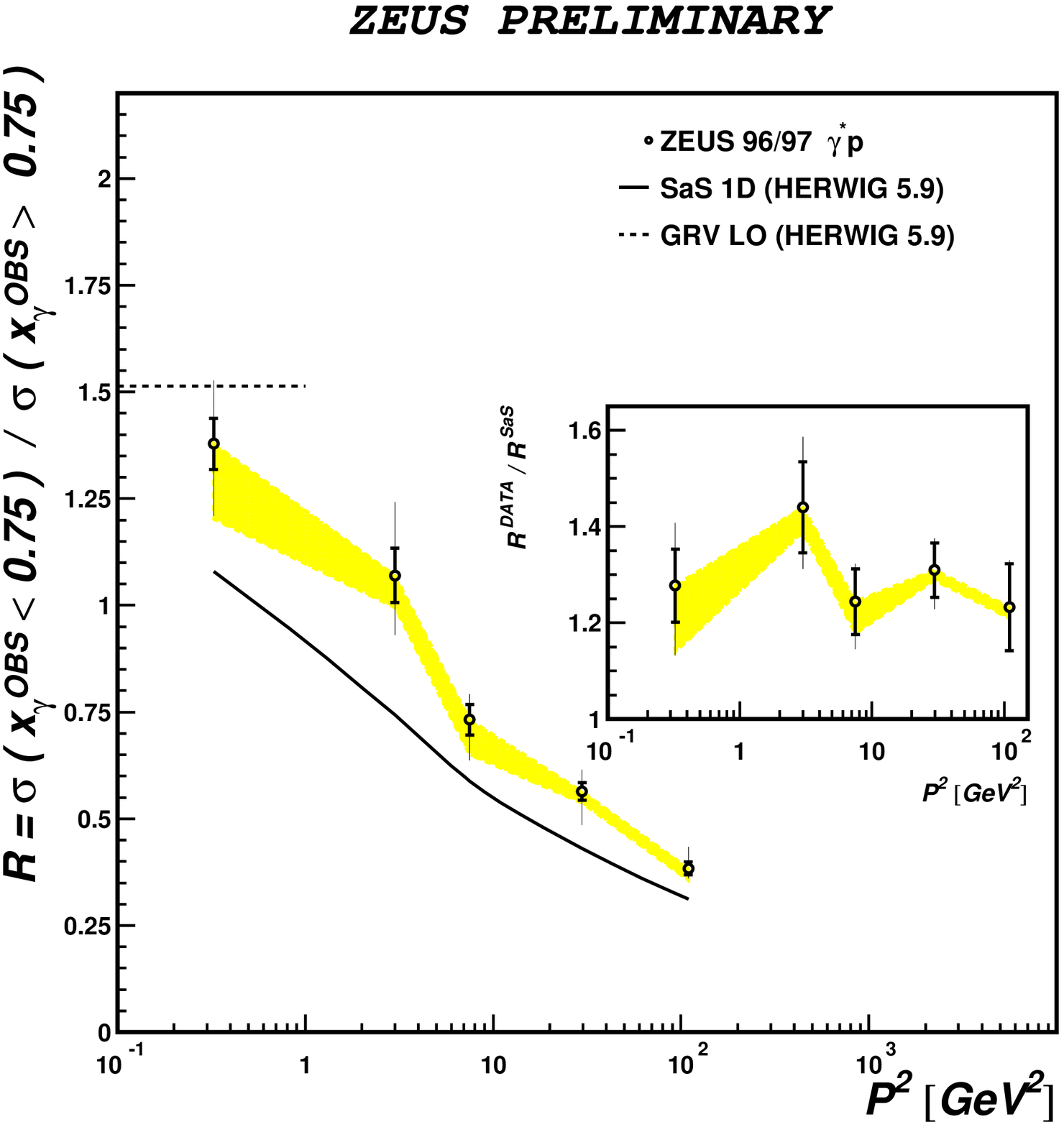,width=8cm}}
\put (0.0,1.5){\small (a)}
\put (8.0,1.5){\small (b)}
\end{picture}
\end{center}
\vspace{-1.5cm}
\caption{\label{fig6}
(a) The measured dijet $\sxo$ (black dots) in the $\gamma^*p$ frame.
(b) The ratio of dijet cross sections, $\sigma(\xo<0.75)/\sigma(\xo>0.75)$, 
as a function of $\p2$ (black dots). Monte Carlo calculations using HERWIG are
shown for comparison.}
\end{figure*}

Figure \ref{fig6}a shows the measurements of the dijet cross section as a
function of $\xo$ in the $\gamma^*p$ frame \cite{gpf}. The measurements have
been performed in the kinematic region given by $0.2<y<0.55$ and
$0.1<\p2<10^4$ \g2. The events are required to have at least two jets with
$E_{T,\gamma^*p}^{jet1(2)}>7.5 (6.5)$ GeV and $-3<\eta_{\gamma^*p}^{jet1,2}<0$.
Also for these measurements the shape of the measured cross section
changes with increasing $\p2$: the cross section for low $\xo$ values falls
more rapidly with increasing $\p2$ than at high $\xo$ values.

The predictions of HERWIG based on the SaS 1D PDFs (see figure \ref{fig6}a)
give a good description of the data in the high $\p2$ region but fail in the
intermediate $\p2$ region. A resolved photon component is needed to describe
the data up to $\p2\sim 49$ \g2. Above this value, the HERWIG prediction is
dominated by the direct component and describes the data well. Therefore,
for $\p2\geq(\etjet)^2$, direct processes alone are able to describe the data.

The ratio of the cross section for $\xo<0.75$ and $\xo>0.75$ also falls steeply
in the $\gamma^*p$ frame with increasing $\p2$ (figure \ref{fig6}b). The
prediction of SaS 1D shows a decrease with $\p2$ but lies below the data in the
whole $\p2$ range. The ratio of the data to the SaS 1D prediction (inset in
figure \ref{fig6}b) has a constant value of $\sim 1.3$. This indicates that
the resolved component suppression included in these PDFs is in agreement with
the data, but they underestimate the fraction of resolved component by
$\sim 30\%$.

\section*{SUMMARY AND CONCLUSIONS}

Dijet cross sections as a function of $\xo$ have been measured in the
laboratory and $\gamma^*p$ frames in the kinematic region given by $0.2<y<0.55$
and $0\lesssim\p2<10^4$ \g2\ for jets found using the longitudinally
invariant $\kt$ cluster algorithm in the inclusive mode. The $\xo$ dependence
of the measured cross sections changes with increasing $\p2$: the cross
section for low $\xo$ values decreases faster than at high $\xo$ values. The
predictions of HERWIG based on the SaS 1D photon PDFs describe the data for
$1.5<\p2<10^4$ \g2\ but fail in the region $0.1<\p2<0.55$ \g2. A resolved
photon component is needed to describe the data up to $\p2\sim 49$ \g2, i.e.
where the virtual photon is probed at a scale comparable to the hard
interaction scale $(\etjet)^2\sim(7\ {\rm GeV})^2$. The ratio of the dijet
cross section $\sigma(\xo<0.75)/\sigma(\xo>0.75)$ decreases with $\p2$; the
predicted $\p2$ dependence of the ratio agrees with the data, but it
underestimates the fraction of the resolved component by $\sim 30\%$. This
result can be interpreted in terms of a resolved photon component that is
suppressed as the photon virtuality increases.

\vspace{0.5cm}
{\bf Acknowledgements.} I would like to thank my colleagues from ZEUS for
their help in preparing this report and the organisers of the conference for
providing a warm atmosphere and hospitality.


\end{document}